\documentclass[prd,twocolumn,preprintnumbers,amsmath,amssymb,nofootinbib]{revtex4}

\usepackage{amsmath}
\usepackage{graphicx}
\usepackage{xspace}
\usepackage{color}
\usepackage{url}
\usepackage{overpic}
\usepackage[utf8]{inputenc}
\usepackage{epstopdf}
\usepackage{hyperref}

\newcommand{\bea}{\begin{eqnarray}}
\newcommand{\eea}{\end{eqnarray}}

\newcommand{\eq}[1]{Eq.~\eqref{#1}}

\begin{document}
\preprint{CERN-TH-2020-134, PSI-PR-20-12, ZU-TH 28/20}

\title{Correlating \begin{boldmath}$h\to\mu^+\mu^-$\end{boldmath} to the Anomalous Magnetic Moment of the Muon \\ via Leptoquarks}

\author{Andreas Crivellin}
\email{andreas.crivellin@cern.ch}
\affiliation{CERN Theory Division, CH--1211 Geneva 23, Switzerland}
\affiliation{Physik-Institut, Universit\"at Z\"urich,
	Winterthurerstrasse 190, CH-8057 Z\"urich, Switzerland}
\affiliation{Paul Scherrer Institut, CH--5232 Villigen PSI, Switzerland}

\author{Dario M\"uller}
\email{dario.mueller@psi.ch}
\affiliation{Physik-Institut, Universit\"at Z\"urich,
	Winterthurerstrasse 190, CH-8057 Z\"urich, Switzerland}
\affiliation{Paul Scherrer Institut, CH--5232 Villigen PSI, Switzerland}

\author{Francesco Saturnino}
\email{saturnino@itp.unibe.ch}
\affiliation{Albert Einstein Center for Fundamental Physics, Institute
	for Theoretical Physics,\\ University of Bern, CH-3012 Bern,
	Switzerland}

\begin{abstract}
Recently, both ATLAS and CMS measured the decay $h\to\mu^+\mu^-$, finding a signal strength with respect to the Standard Model expectation of $1.2\pm0.6$ and $1.19\substack{+0.41+0.17\\-0.39 -0.16}$, respectively. This provides, for the first time, evidence that the Standard Model Higgs couples to second generation fermions. This measurement is particularly interesting in the context of the intriguing hints for lepton flavor universality violation, accumulated within recent years, as new physics explanations could also be tested in the $h\to\mu^{+}\mu^{-}$ decay mode. Leptoquarks are prime candidates to account for the flavor anomalies. In particular, they can provide the necessary chiral enhancement (by a factor $m_t/m_\mu$) to address $a_\mu$ with TeV scale new physics. In this letter we point out that such explanations of $a_\mu$ also lead to enhanced effects in $h\to\mu^+\mu^-$ and we examine the correlations between $h\to\mu^+\mu^-$ and $a_\mu$ within leptoquark models. We find that the effect in the branching ratio of $h\to\mu^+\mu^-$ ranges from several percent up to a factor three, if one aims at accounting for $a_\mu$ at the $2\,\sigma$ level. Hence, the new ATLAS and CMS measurements already provide important constraints on the parameter space, rule out specific  $a_\mu$ explanations and will be very important to test the flavor anomalies in the future. 
\end{abstract}

\maketitle

\section{Introduction}

The Large Hadron Collider (LHC) at CERN confirmed the predictions of the Standard Model (SM) of particle physics by discovering the Brout-Englert-Higgs boson~\cite{Aad:2012tfa,Chatrchyan:2012xdj} in 2012. However, until now, high energy searches did not discover any particles beyond the ones present in the SM. Therefore, great hopes of finding new physics (NP) rest on low energy precision physics where flavor experiments have accumulated intriguing hints for physics beyond the SM within the recent years, most prominently in $b\to s\ell^+\ell^-$ data~\cite{Aaij:2017vbb,Aaij:2019wad,Aaij:2020nrf}, $b\to c\tau\nu$ transitions~\cite{Lees:2012xj,Aaij:2017uff,Abdesselam:2019dgh} and the anomalous magnetic moment (AMM) of the muon ($a_\mu=(g-2)_{\mu}/2$)~\cite{Bennett:2006fi,Mohr:2015ccw,Abi:2021gix}. Interestingly, these hints for NP fall into a common pattern: they can be considered as signs of lepton flavor universality violation (LFUV)~\footnote{Recently, it has been pointed out that also the Cabibbo Angle Anomaly can be interpreted as a sign of LFUV~\cite{Coutinho:2019aiy,Crivellin:2020lzu}.}, which is respected by the SM gauge interactions and is only broken by the Higgs Yukawa couplings.

Among these anomalies, $a_\mu$, which displays a $4.2\,\sigma$ deviation from the SM prediction~\cite{Aoyama:2020ynm}, is most closely related to Higgs interactions as it is a chirality changing observable. I.e. it involves a chirality flip and therefore a violation of $SU(2)_L$ is required to obtain a non-zero contribution. Furthermore, the required NP effect to explain $a_\mu$ is of the order of the electroweak (EW) SM contribution and TeV scale solutions need an enhancement mechanism, called chiral enhancement, to be able to account for the deviation (see e.g. Ref.~\cite{Crivellin:2018qmi} for a recent discussion). Obviously, also $h\to\mu^+\mu^-$ is a chirality changing process and any enhanced effect in $a_\mu$ should also result in an enhanced effect here~\footnote{Correlations between $a_\mu$ and $h\to\mu^+\mu^-$ were considered in the EFT in Ref.~\cite{Feruglio:2018fxo} and in the context of vector-like leptons (see Ref.~\cite{Crivellin:2020ebi} for a recent global analysis) in Ref.~\cite{Kannike:2011ng,Dermisek:2013gta,Dermisek:2014cia,Crivellin:2018qmi}.}. Recently, both ATLAS and CMS measured $h\to\mu^+\mu^-$, finding a signal strength w.r.t. the SM expectation of $1.2\pm0.6$ \cite{Aad:2020xfq} and $1.19^{+0.41+0.17}_{-0.39 -0.16}$ \cite{CMS:2020eni}, respectively. 

The mechanism of chiral enhancement, necessary to explain $a_\mu$, has been well studied (see Ref.~\cite{Crivellin:2018qmi} for a recent account). 
Here leptoquarks (LQs) are particularly interesting since they can give rise to an enhancement factor of {$m_t/m_\mu \approx 1700$}~\cite{Djouadi:1989md, Chakraverty:2001yg,Cheung:2001ip,Bauer:2015knc,Popov:2016fzr,Chen:2016dip,Biggio:2016wyy,Davidson:1993qk,Couture:1995he,Mahanta:2001yc,Queiroz:2014pra,ColuccioLeskow:2016dox,Chen:2017hir,Das:2016vkr,Crivellin:2017zlb,Cai:2017wry,Crivellin:2018qmi,Kowalska:2018ulj,Mandal:2019gff,Dorsner:2019itg,Crivellin:2019dwb,DelleRose:2020qak,Saad:2020ihm,Bigaran:2020jil,Dorsner:2020aaz}, allowing for a TeV scale explanation with perturbative couplings that are not in conflict with direct LHC searches. In fact, there are only two LQs, out of the 10 possible representations~\cite{Buchmuller:1986zs}, that can yield this enhancement: the scalar LQ $SU(2)_L$ singlet ($S_1$) and the scalar LQ $SU(2)_L$ doublet ($S_2$) with hypercharge $-2/3$ and $-7/3$, respectively. In addition, there is the possibility that $S_1$ mixes with the $SU(2)_L$ triplet LQ $S_3$, where $S_1$ only couples to right-handed fermions~\cite{Dorsner:2019itg}.

Furthermore, LQs are also well motivated by the hints for LFUV in semi-leptonic $B$ decays, both in $b\to s\mu^+\mu^-$~\cite{Aaij:2017vbb,Aaij:2019wad,Aaij:2020nrf} and $b\to c\tau\nu$ data~\cite{Lees:2012xj,Aaij:2017uff,Abdesselam:2019dgh}, which deviate from the SM with up to $\approx 6\,\sigma$~\cite{Alguero:2019ptt,Aebischer:2019mlg,Ciuchini:2019usw,Arbey:2019duh} and $\approx 3\,\sigma$~\cite{Amhis:2019ckw,Murgui:2019czp,Shi:2019gxi,Blanke:2019qrx,Kumbhakar:2019avh}, respectively. Here possible solutions include again $S_1$~\cite{Fajfer:2012jt, Deshpande:2012rr, Sakaki:2013bfa, Freytsis:2015qca, Bauer:2015knc, Li:2016vvp, Zhu:2016xdg, Popov:2016fzr, Deshpand:2016cpw, Becirevic:2016oho, Cai:2017wry, Buttazzo:2017ixm, Altmannshofer:2017poe, Kamali:2018fhr, Azatov:2018knx, Wei:2018vmk, Angelescu:2018tyl, Kim:2018oih, Crivellin:2019qnh, Yan:2019hpm}, $S_2$~\cite{Tanaka:2012nw, Dorsner:2013tla, Sakaki:2013bfa, Sahoo:2015wya, Chen:2016dip, Dey:2017ede, Becirevic:2017jtw, Chauhan:2017ndd, Becirevic:2018afm, Popov:2019tyc} and $S_3$~\cite{Fajfer:2015ycq, Varzielas:2015iva, Bhattacharya:2016mcc, Buttazzo:2017ixm, Barbieri:2015yvd, Kumar:2018kmr, deMedeirosVarzielas:2019lgb, Bernigaud:2019bfy}, where $S_1$ and $S_3$ together can provide a common explanation of the $B$ anomalies and the AMM of the muon~\cite{Crivellin:2017zlb,Buttazzo:2017ixm,Marzocca:2018wcf, Bigaran:2019bqv,Crivellin:2019dwb}. We take this as a motivation to study these correlations for the LQs which can generate $m_t/m_\mu$ enhanced effects by considering three scenarios: 1) $S_1$ only, 2) $S_2$ only, 3) $S_1+S_3$ where $S_1$ only couples to right-handed fermions. Note that these are the only scenarios which can give rise to the desired $m_t/m_\mu$ enhanced effect.

\begin{figure}[t!]
	\centering
	\begin{overpic}[scale=.55,,tics=10]
		{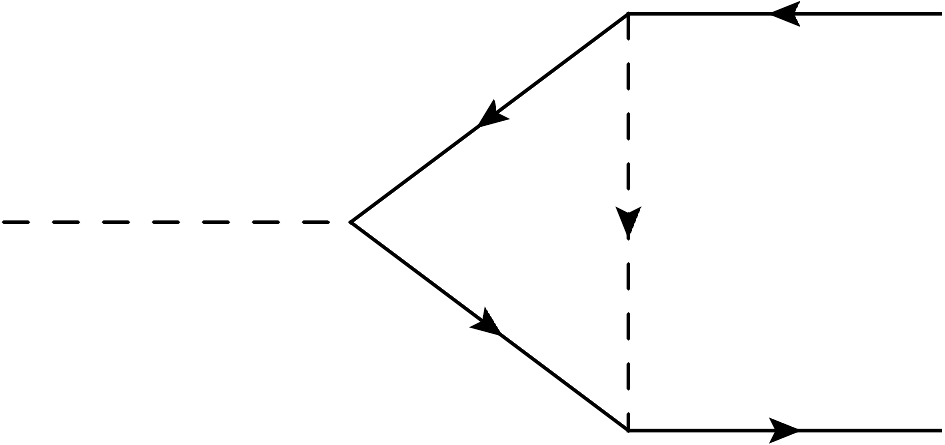}
		\put(5,27){$h$}
		\put(90,39){$\mu$}
		\put(90,6){$\mu$}
		\put(42,36){$t^{(c)}$}
		\put(42,5){$t^{(c)}$}
		\put(70,22){$S_{i}$}
	\end{overpic}
	\\
	\vspace{5mm}
	\begin{overpic}[scale=.55,,tics=10]
		{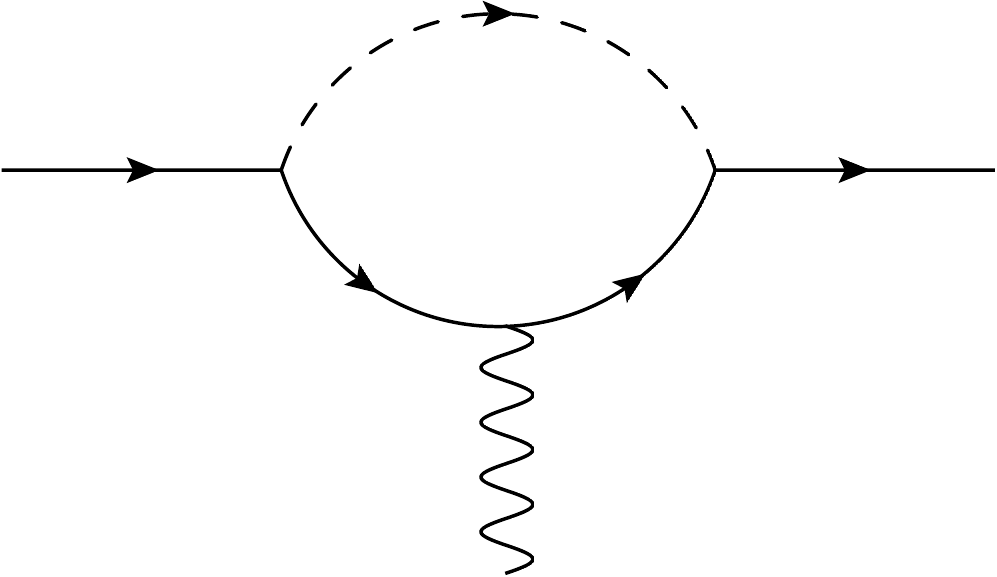}
		\put(5,45){$\mu$}
		\put(90,45){$\mu$}
		\put(57,5){$\gamma$}
		\put(48,48){$S_{i}$}
		\put(27,24){$t^{(c)}$}
		\put(66,24){$t^{(c)}$}
	\end{overpic}
	\caption{Sample Feynman diagrams which contribute to $h\to\mu^{+}\mu^-$ (top) and the AMM of the muon (bottom). In addition, we have to include the diagrams where the Higgs and photon couple to the LQ, as well as self-energy diagrams.}
	\label{FeynmanDiagrams}
\end{figure}
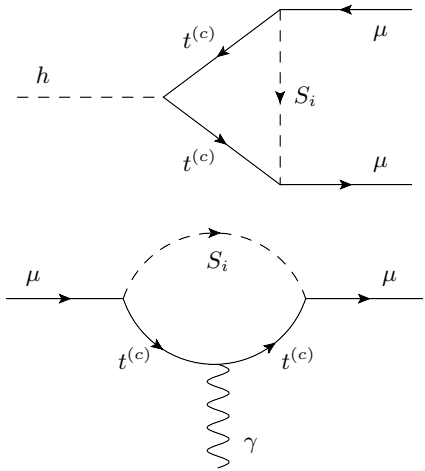

\section{Setup and Observables}

The most precise measurements of the anomalous magnetic moment (AMM) of the muon ($a_\mu=(g-2)_{\mu}/2$) has been achieved by the E821 experiment at Brookhaven~\cite{Bennett:2006fi,Mohr:2015ccw} and recently be the g-2 experiment at Fermilab~\cite{Abi:2021gix}, which differs from the SM prediction by
\begin{equation}
\label{Delta_amu}
\delta a_\mu=a_\mu^{\rm{exp}} - a_\mu^{\rm{SM}} = (251 \pm 59) \times 10^{-11} \,,
\end{equation}
corresponding to a $4.2\,\sigma$ deviation~\cite{Aoyama:2020ynm}\footnote{This result is based on Refs.~\cite{Aoyama:2012wk,Aoyama:2019ryr,Czarnecki:2002nt,Gnendiger:2013pva,Davier:2017zfy,Keshavarzi:2018mgv,Colangelo:2018mtw,Hoferichter:2019gzf,Davier:2019can,Keshavarzi:2019abf,Kurz:2014wya,Melnikov:2003xd,Masjuan:2017tvw,Colangelo:2017fiz,Hoferichter:2018kwz,Gerardin:2019vio,Bijnens:2019ghy,Colangelo:2019uex,Blum:2019ugy,Colangelo:2014qya}. The recent lattice result of the Budapest-Marseilles-Wuppertal collaboration (BMWc) for the hadronic vacuum polarization (HVP)~\cite{Borsanyi:2020mff} on the other hand is not included. This result would render the SM prediction of $a_\mu$ compatible with experiment. However, the BMWc results are in tension with the HVP determined from $e^+e^-\to$ hadrons data~\cite{Davier:2017zfy,Keshavarzi:2018mgv,Colangelo:2018mtw,Hoferichter:2019gzf,Davier:2019can,Keshavarzi:2019abf}. Furthermore, the HVP also enters the global EW fit~\cite{Passera:2008jk}, whose (indirect) determination is below the BMWc result~\cite{Haller:2018nnx}. Therefore, the BMWc determination of the HVP would increase tension in EW fit~\cite{Crivellin:2020zul,Keshavarzi:2020bfy} and we opted for using the community consensus of Ref.~\cite{Aoyama:2020ynm}.}. Therefore, it is very interesting to investigate if and how this discrepancy can be explained by physics beyond the SM.

\begin{table}
\begin{equation}
\renewcommand{\arraystretch}{2}
\begin{tabular}{c|c|c}
& $\mathcal{G}_{\text{SM}}$ & $\mathcal{L}_{q\ell}$\\
\hline
$S_1$ & $\bigg(3,1,-\dfrac{2}{3}\bigg)$ & $\left(\lambda_{fj}^{R}\,\bar{u}^c_f\ell_{j}+\lambda_{fj}^{L}\,\bar{Q}_{f}^{\,c}i\tau_{2}L_{j}\right) S_{1}^{\dagger}+\text{h.c.}$\\
$S_{2}$ & $\bigg(3,2,\dfrac{7}{3}\bigg)$ & $\gamma_{fj}^{RL}\,\bar{u}_{f}S_{2}^{T}i\tau_{2}L_{j}+\gamma_{fj}^{LR}\,\bar{Q}_f\ell_j S_{2}+\text{h.c.}$\\
$S_{3}$ & $\bigg(3,3,-\dfrac{2}{3}\bigg)$ & $\kappa_{fj}^{}\,\bar{Q}^{\,c}_{f}i\tau_{2}\left(\tau\cdot S_{3}\right)^{\dagger}L_{j}+\text{h.c.}$
\end{tabular}\nonumber
\end{equation}
\caption{Scalar LQ representations together with their couplings to quarks and leptons, generating the desired $m_t/m_\mu$ enhanced effect in the AMM of the muon. Here  $\mathcal{G}_{\text{SM}}$ refers to the SM gauge group $SU(3)_c\times SU(2)_L\times U(1)_Y$,  $L$ ($Q$) is the lepton (quark) $SU(2)_{L}$ doublet, $u$ ($\ell$) the up-type quark (lepton) singlet and $c$ refers to charge conjugation. Furthermore, $j$ and $f$ are flavor indices and $\tau_{k}$ the Pauli matrices.}
\label{LQrep}
\end{table}

As we motivated in the introduction, we will focus on the three scalar LQs $S_1$, $S_2$ and $S_3$ for explaing $a_\mu$. These representations couple to fermions as given in Table~\ref{LQrep}\footnote{Note that ``pure'' LQs with couplings only to one quark and one lepton do not give rise to proton decays at any perturbative order. The reason for this is that di-quark couplings are necessary in order to break baryon and/or lepton number which is otherwise an unbroken symmetry forbidding proton decay (see Ref.~\cite{Dorsner:2012nq} for a recent detailed discussion).}. Since we are in the following only interested in muon couplings to third generation quarks, we define
$\lambda _R^{} \equiv \lambda _{32}^R$, $\lambda _L^{} \equiv \lambda _{32}^L$, $\gamma _{LR}^{} \equiv \gamma _{32}^{LR},$ $\gamma _{RL}^{} \equiv \gamma _{32}^{RL}$, $\kappa  = {\kappa _{32}}$. 

In addition to the gauge interactions, which are determined by the representation under the SM gauge group, LQ can couple to the SM Higgs~\cite{Hirsch:1996qy}
\begin{align}
{{\cal L}_H} &= {Y_{13}}S_1^\dag \left( {{H^\dag }\left( {\tau \cdot{S_3}} \right)H} \right) + {\rm{h}}{\rm{.c}}{\rm{.}} \label{eq:LQ_mixing}\\
&- {Y_{22}}{\left( {Hi{\tau _2}{S_2}} \right)^\dag }\left( {Hi{\tau _2}{S_2}} \right) - \sum\limits_{k = 1}^3 ( m_k^2 + {Y_k}{H^\dag }H)S_k^\dag {S_k}
\nonumber
\end{align}
Here $m_k^2$ are the $SU(2)_L$ invariant bi-linear masses of the LQs. After $SU(2)_L$ breaking, the term $Y_{13}$ generates off-diagonal elements in the LQ mass matrices and one has to diagonalize them through unitary transformations in order to arrive at the physical basis. Therefore, non-zero values of $Y_{13}$ are necessary to generate $m_t/m_\mu$ enhanced effects in scenario 3). $Y_1$ and $Y_{2,22}$ are phenomenologically relevant for $h\to\mu^+\mu^-$ in scenario 1) and 2), respectively, but not necessary for an $m_t/m_\mu$ enhancement.

\begin{figure*}
	\centering
	\includegraphics[width=0.54\textwidth]{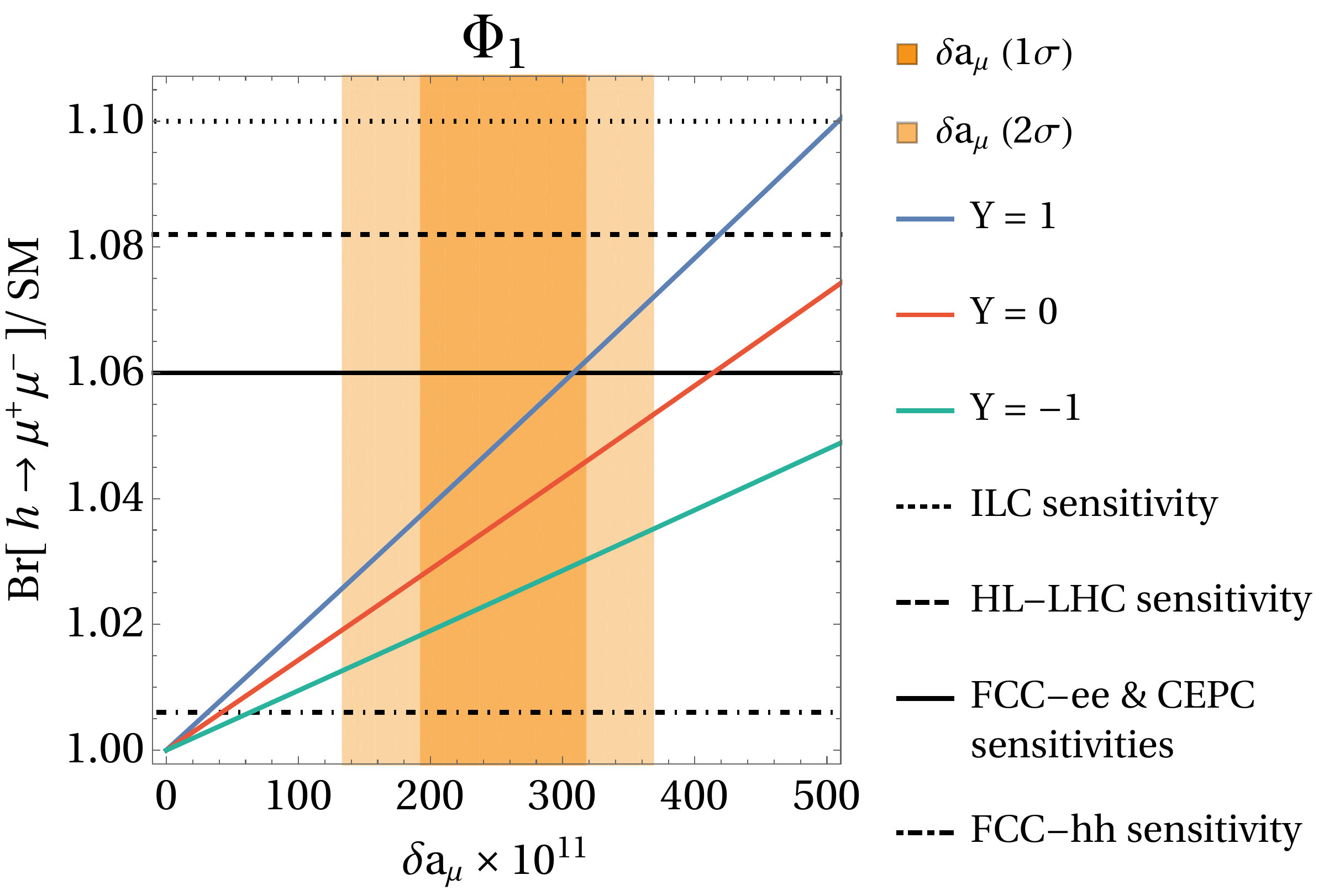}
	\includegraphics[width=0.352\textwidth]{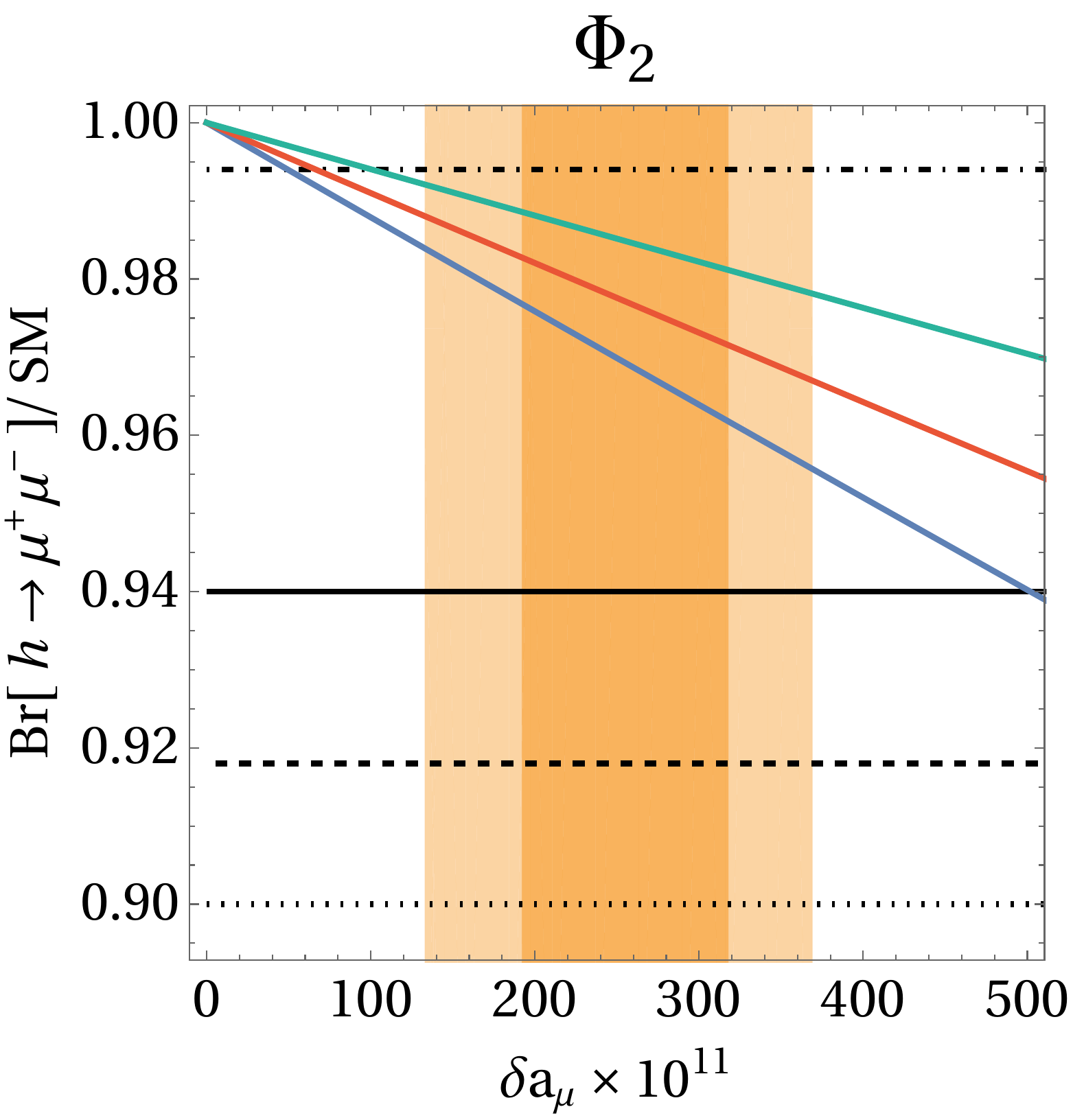}
	\caption{Correlations between the ${\rm Br}[h\to\mu^+\mu^-]$, normalized to its SM value, and the NP contribution in the AMM of the muon $\delta a_\mu$ for scenario 1) (left) and scenario 2) (right) with $m_{1,2}=1.5\,$TeV. The predictions for different values of the LQ couplings to the Higgs are shown, where for scenario 1) $Y=Y_1$ while in scenario 2) $Y=Y_2+Y_{22}$. Even though the current ATLAS and CMS results are not yet constraining for these models, sizeable effects are predicted, which can be tested at future colliders. Furthermore, scenario 1) yields a constructive effect in $h\to\mu^+\mu^-$ while the one in scenario 2) is destructive such that they can be clearly distinguished with increasing experimental precision. }\label{S1S2}
\end{figure*}

Now we can calculate the effects in $a_\mu$ and $h\to\mu^+\mu^-$~\footnote{Correlations between the related modes $\tau\to\mu\gamma$ and $h\to\tau\mu$ were studied in Refs.~\cite{Dorsner:2015mja,Cheung:2015yga,Baek:2015mea} in the context of LQs} for which sample diagrams are shown in Fig.~\ref{FeynmanDiagrams}. In both cases we have on-shell kinematics. For $a_\mu$ the self-energies can simply be taken into account via the Lehmann-Symanzik-Zimmermann formalism and no renormalization is necessary. This is however required for $h\to\mu^+\mu^-$ in order to express the result in terms of the physical muon mass. Here, the effective Yukawa coupling, which enters $h\to\mu^+\mu^-$, is given by
\begin{equation}
Y_\mu ^{{\rm{eff}}} = \frac{{m_\mu ^{} - \Sigma _{\mu \mu }^{LR}}}{v} + \Lambda _{\mu \mu }^{LR}\,,
\end{equation}
where $\Lambda _{\mu \mu }^{LR}$ is the genuine vertex correction shown in Fig.~\ref{FeynmanDiagrams} and $\Sigma _{\mu \mu }^{LR}$ is the chirality changing part of the muon self-energy. In these conventions $-i\Sigma _{\mu \mu }^{LR}P_R$ equals the expression of the Feynman diagram for the self-energy. Note that $Y_\mu ^{{\rm{eff}}}$ is finite without introducing a counter-term. For $a_\mu$ we expand in the muon mass and external momenta up to the first non-vanishing order, while in $h\to\mu^+\mu^-$ external momenta can be set to zero from the outset but we expand in $m_h^2/m_{1,2,3}^2$. The resulting amplitudes can be further simplified by expanding the LQ mixing matrices and mass eigenvalues in $v^2/m_{1,2,3}^2$ and the loop functions in $m_h^2/m_t^2$, which gives a very precise numerical approximation, resulting in
\begin{widetext}
	\begin{align}
	\dfrac{{{\rm{Br}}\left[ {h \to \mu^{+} \mu^{-} } \right]}}{{{\rm{Br}}{{\left[ {h \to \mu^{+} \mu^{-} } \right]}_{{\rm{SM}}}}}} \approx \Bigg| 
	1 + \dfrac{m_t}{m_\mu }\dfrac{N_c}{8\pi^2}\bigg[ \dfrac{\lambda _R^*\lambda_L}{m_1^2}\left( \dfrac{m_t^2}{8}\mathcal{J}\!\left( {\dfrac{{{m_h^2}}}{{m_t^2}},\dfrac{{m_t^2}}{{m_1^2}}} \right) + v^{2}Y_1 \right)+v^{2}\lambda_R^{*}\kappa Y_{13} {\dfrac{{\log \left( {m_3^2/m_1^2} \right)}}{{m_3^2-m_1^2}}}\nonumber\\
	\qquad\qquad\qquad+ \dfrac{\gamma_{LR}^*\gamma _{RL}}{m_{2}^2}\left( \dfrac{m_t^2}{8}\mathcal {J}\!\left( {\dfrac{{{m_h^2}}}{{m_t^2}},\dfrac{{m_t^2}}{{m_2^2}}} \right) + v^{2}{(Y_2+ Y_{22})} \right)\bigg]
	\Bigg|^2\,,\\
	{\delta a_\mu }\approx \frac{{{m_\mu }}}{{4{\pi ^2}}}\frac{{{N_c}{m_t}}}{{12}}{\rm{Re}}\left[ 
	\dfrac{\gamma_{LR}^{}\gamma _{RL}^*}{m_{2}^2} { {\cal E}_{1}\!\left( {\frac{{m_t^2}}{{m_2^2}}} \right)} - \frac{\lambda_R}{m_{1}^2} \left( {\lambda _L^* {{\cal E}_{2}\!\left( {\frac{{m_t^2}}{{m_1^2}}} \right)} + \kappa {Y_{13}}\frac{{{v^2}}}{{m_3^2}}{\cal E}_{3}\!\left( {\frac{{m_1^2}}{{m_3^2}}},\frac{{m_t^2}}{{m_3^2}} \right)} \right)\right]\,,
\label{hmumuamuFormula}
	\end{align}
\end{widetext}
with the loop functions given by
\begin{align}
{\cal J}\left( x,y \right) &= 2\left( {x - 4} \right)\log (y) - 8 + \frac{{13}}{3}x{\mkern 1mu} \,,
\end{align}
\begin{align}
\begin{aligned}
{\cal E}_{1}(x)&=1+4\,\log(x)\,,\;\;
{\cal E}_{2}(x)=7+4\,\log(x)\,,\\
{\cal E}_{3}( x,y ) &= {\cal E}_{2}(y) + \frac{4\,{\log (x)}}{{x - 1}} \,.
\end{aligned}
\end{align}
We only considered the $m_t$ enhanced effects and neglected small CKM rotations, which in principle appear after EW symmetry breaking. As anticipated, in \eq{hmumuamuFormula} one can see that scenario 3) only contributes if $Y_{13}$ is non-zero. Furthermore, since in this scenario $a_\mu$ has a relative suppression of $v^2/m_{1,3}^2$ with respect to $h\to\mu^+\mu^-$, one expects here the largest effects in Higgs decays. In principle also $Y_1$, $Y_2$ and $Y_{22}$ enter in \eq{hmumuamuFormula}. However, their effect is sub-leading as it is suppressed by $v^2/m_{1,2}^2$.

\subsection{Effective Field Theory}

In the SM effective field theory (SMEFT), which is realized above the EW breaking scale and therefore explicitly $SU(2)_L$ invariant, there are only two chirality flipping 4-fermion operators~\cite{Grzadkowski:2010es} which can give rise to $m_t$ enhanced effects in $a_\mu$ and $h\to\mu^+\mu^-$ via re-normalization group evolution (RGE) effects:
\begin{align}
\begin{aligned}
Q_{{\rm{\ell equ }}}^{(1)} &= \left( {\bar \ell _2^a{e_2}} \right){\varepsilon _{ab}}\left( {\bar q_3^b{u_3}} \right)\,,\\
Q_{{\rm{\ell  equ }}}^{(3)} &= \left( {\bar \ell _2^a{\sigma _{\mu \nu }}{e_2}} \right){\varepsilon _{ab}}\left( {\bar q_3^b{\sigma ^{\mu \nu }}{u_3}} \right)\,.
\end{aligned}
\end{align}
Importantly, while both operators mix at order $\alpha_{(s)}$ with each other, only the second operators mixes (directly) into the magnetic operator~\cite{Jenkins:2013zja,Jenkins:2013wua,Alonso:2013hga}
\begin{equation}
\begin{aligned}
{Q_{eB}} &= {{\bar \ell }_2}{\sigma^{\mu \nu }}{e_2}H{B_{\mu \nu }}\,,\\
{Q_{eW}} &= {{\bar \ell }_2}{\sigma^{\mu \nu }}{e_2}{\tau ^I}HW_{\mu \nu }^I\,,
\end{aligned}
\end{equation}
giving rise to the AMM of the muon after EW symmetry breaking\footnote{Note that LQs are the only renormalizable extensions of the SM that can generate these operator at tree-level~\cite{deBlas:2017xtg}.}.  Furthermore, as $Q_{{\rm{\ell equ }}}^{(1)}$ mixes into ${Q_{e\varphi }} =  {{H^\dag }H}  {{{\bar \ell }_2}{e_2}H}$ (generating modified Higgs couplings to muons) it is clear that a UV complete (or at least simplified) model is necessary to correlate $a_\mu$ to $h\to\mu^+\mu^-$.

The EFT approach is beneficial in our LQ setup since it allows for the inclusion of RGE effects, as recently done in Ref.~\cite{Aebischer:2021uvt}. In a first step, the LQ model is matched on the SMEFT (at the LQ scale), giving tree-level effects in $C_{{\rm{\ell equ }}}^{(1,3)}$~\cite{Alonso:2015sja} and a loop effect in ${Q_{eB}}$ and 
${Q_{eW}}$~\cite{Gherardi:2020det}. Then the SMEFT is used to evolve the Wilson coefficients of these operators to the weak scale where the EW gauge bosons, the Higgs and the top quark are integrated out~\cite{Crivellin:2013hpa,Dekens:2019ept,Hurth:2019ula}. Next, the magnetic operator of the muon is evolved to the muon scale~\cite{Crivellin:2017rmk,Aebischer:2017gaw} where the AMM is measured. Ref.~\cite{Aebischer:2021uvt} finds a reduction of $a_\mu$ by $\approx20\%-30\%$ compared to the leading order estimate of LQ masses between $1$--$10\,$TeV. Furthermore, as $C_{{\rm{lequ }}}^{(1)}$ is enhanced by $\approx5\%-10\%$ by the running from the LQ scale to the EW scale~\cite{Aebischer:2018bkb}, this leads to an important enhancement of $50\%-70\%$ of the prediction for ${\rm Br}[h\to\mu^+\mu^-]$ w.r.t the leading order calculation. To be conservative, we will use $50\%$ in our following phenomenological analysis.

\section{Phenomenology}
\label{pheno}

Let us now study the correlations between $a_\mu$ and $h\to\mu^+\mu^-$ in our three scenarios with $m_t$-enhanced contributions. First, we consider scenario 1) and 2) where $S_1$ and $S_2$ give separately rise to $m_t$-enhanced effects in $a_\mu$ and $h\to\mu^+\mu^-$. Since both processes involve the same product of couplings to SM fermions, the correlation depends only weakly via a logarithm on $m_t^2/m_{1,2}^2$. However, there is a dependence on $Y_1$ and $Y_{22}+Y_2$ which breaks the direct correlation but cannot change the sign of the effect for order one couplings. This can be seen in Fig.~\ref{S1S2}, where the correlations are depicted for $m_{1,2}=1.5$ TeV, respecting LHC bounds~\cite{Sirunyan:2018ryt,Diaz:2017lit,Aaboud:2016qeg}. The predicted effect is not large enough such that the current ATLAS and CMS measurements are sensitive to it. However, note that it is still sizeable due to the $m_t$ enhancement and therefore detectable at future colliders where the ILC~\cite{Behnke:2013lya}, the HL-LHC~\cite{ApollinariG.:2017ojx}, the FCC-ee~\cite{Abada:2019zxq}, CEPC~\cite{An:2018dwb} or the FCC-hh~\cite{Benedikt:2018csr} aim at a precision of approximately 10\%, 8\%, 6\% and below 1\%, respectively. Furthermore, the effect in ${\rm Br}[h\to\mu^+\mu^-]$ in scenario 1) is necessarily constructive while in scenario 2) it is destructive, such that in the future a LQ explanation of $a_\mu$ by $S_1$ could be clearly distinguished from the one involving $S_2$. 

\begin{figure*}
	\centering
	\includegraphics[width=0.52\textwidth]{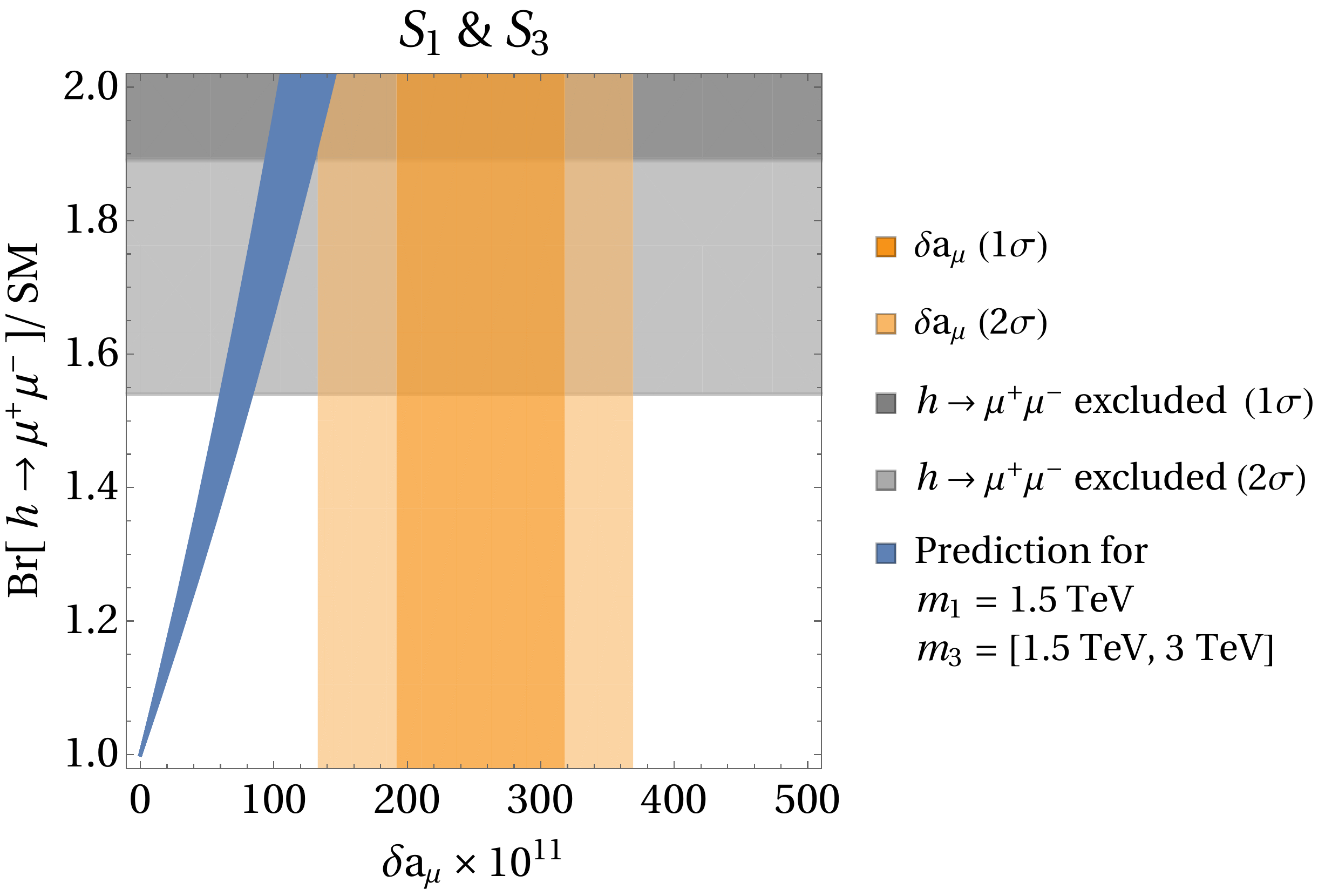}
	\caption{Correlations between the NP contribution to the AMM of the muon ($\delta a_\mu$) and ${\rm Br}[h\to\mu^+\mu^-]$, normalized to its SM value in scenario 3). This correlation depends to a good approximation only on the ratio $m_1/m_3$. As the effect is symmetric in $m_1$ and $m_3$, we fix one mass to $1.5\,$TeV and obtain the dark-blue band by varying the other mass between $1.5\,$TeV and $3\,$TeV. The effect in $h\to\mu^+\mu^-$ within the preferred region for $a_\mu$ is necessarily constructive and so large that an explanation is already constrained by the ATLAS and CMS measurements of $h\to\mu^+\mu^-$.}
	\label{S1S3Y13}
\end{figure*}

In scenario 3), where $S_1$ only couples to right-handed fermions, the effect in ${\rm Br}[h\to\mu^+\mu^-]$ is even more pronounced due to the relative suppression of the contribution to $a_\mu$ by $v^2/m_{1,3}^2$, see \eq{hmumuamuFormula}. Furthermore, in this case the correlation between $a_\mu$ and $h\to\mu^+\mu^-$ depends to a good approximation only on the ratio $m_1/m_3$. As the effect is symmetric in $m_1$ and $m_3$ we fix one mass to $1.5$ TeV and obtain the band shown in Fig.~\ref{S1S3Y13} by varying the other mass between $1.5$ and $3$ TeV. The effect in $h\to\mu^+\mu^-$ within the preferred region for $a_\mu$ is necessarily constructive and large enough that an explanation of the central value of $a_\mu$ is already disfavored by the ATLAS and CMS measurements of $h\to\mu^+\mu^-$. Clearly, with more data the LHC will be able to support (disprove) this scenario if it finds a (no) significant enhancement of the $h\to\mu^+\mu^-$ decay, assuming $\delta a_\mu$ is confirmed. This scenario also leads to sizeable effects in $Z\mu\mu$~\cite{Dorsner:2019itg} which are compatible with LEP data~\cite{ALEPH:2005ab}, but could be observed at the ILC~\cite{Behnke:2013lya}, CLIC~\cite{Aicheler:2012bya} or the FCC-ee~\cite{Abada:2019zxq}.

\section{Conclusions}
\label{conclusions}

LQs are prime candidates for an explanation of the intriguing hints of LFUV. As LFUV within the SM only originates from the Higgs, chirality changing observables as the AMM of the muon and, of course, $h\to\mu^+\mu^-$ are especially interesting. In particular, there are three possible LQ scenarios which can address the discrepancy in the AMM of the muon by an $m_t/m_\mu$ enhancement. This also leads to enhanced corrections in $h\to\mu^+\mu^-$, which involve the same coupling structure as the $a_\mu$ contribution. This leads to interesting correlations between $a_\mu$ and $h\to\mu^+\mu^-$, which we study in light of the recent ALTAS and CMS measurements. 

We find that scenario 3), in which $S_1$ only couples to right-handed fermions and mixes after EW symmetry breaking with $S_3$, predicts large constructive effects in $h\to\mu^+\mu^-$ such that the current ATLAS and CMS measurements are already excluding part of the parameter space. In case $\delta a_\mu$ is solely explained by $S_1$ or $S_2$ the effect in  ${\rm  Br}[h\to\mu^+\mu^-]$ is of the order of several percent and therefore detectable at future colliders, in particular at the FCC-hh. Furthermore, while the $S_1$ scenario predicts constructive interference in $h\to\mu^+\mu^-$ for the currently preferred range of $a_\mu$, the $S_2$ scenario predicts destructive interference such that they can be clearly distinguished in the future. 
\medskip

\begin{acknowledgments}
Acknowledgements -- A.C. thanks Martin Hoferichter for useful discussions. The work of A.C. and D.M. supported by a Professorship Grant (PP00P2\_176884) of the Swiss National Science Foundation and the one of F.S. by the Swiss National Science Foundation grant 200020\_175449/1.
\end{acknowledgments}

\bibliography{BIB}

\end{document}